\documentclass{optica-article}

\journal{opticajournal}

\begin{document}

\title{Anomalous diffusion of optical vortices in random wavefields}

\author{Jiaxing Gong, Qi Li, and Jing Wang\authormark{*}}

\address{Department of Biomedical Engineering, College of Life Science and Technology, Huazhong University of Science and Technology, Wuhan 430074, China}

\email{\authormark{*}wang.jing@hust.edu.cn} 



\begin{abstract}
We investigate the dynamic behavior of optical vortices, or phase singularities, in random wavefields and demonstrate the direct experimental observation of the anomalous diffusion of optical vortices. The observed subdiffusion of optical vortices show excellent agreement with the fractional Brownian motion, a Gaussian process. Paradoxically, the vortex displacements are observed exhibiting a non-Gaussian heavy-tailed distribution. We also tune the extent of subdiffusion and non-Gaussianity of optical vortex by varying the viscoelasticity of light scattering media. This complex motion of optical vortices is reminiscent of particles in viscoelastic environments suggesting a vortex tracking based microrheology approach. The fractional Brownian yet non-Gaussian subdiffusion of optical vortices may not only offer insights into the dynamics of phase singularities, but also contribute to the understanding certain general physics, including vortex diffusion in fluids and the decoupling between Brownian and Gaussian.
\end{abstract}

\section{Introduction}
Optical vortices, possessing phase singularities at intensity nulls with optical current flow circulating about, are the most distinctive features of the granular laser speckles arising from the interference of random optical wavefields \cite{Berry2000,Goodman2007}. The accumulated phase change along with a contour surrounding a single vortex is integer times of 2$\pi$, $ q=\frac{1}{2 \pi} \oint \nabla \varphi(x, y) \cdot d \vec{l} $, where the non-zero, signed integer $ q$ is referred to the topological charge of the vortex. Only optical vortices with topological charge $ \pm $1 can stabely exist in random wave fields \cite{Freund1994a}. All equiphase lines of speckle fields converge to the phase singularities. The optical vortices connected by equiphase lines form a network spreading over the whole speckle field and determine the skeleton structure of random wave fields, as shown in Fig. \ref{fig:Vortex}(a). The geometrical and statistical properties of optical vortices, such as the eccentricity of elliptical intensity contours about vortex cores \cite{Berry2000,Wang2005-2}, and the spatial distributions of optical vortices \cite{Shvartsman1994,Freund2015,DeAngelis2016}, have been intensively studied to advance the understanding of random wave fields which have been a topic of great interest for decades.

Since the evolution of dynamic speckle fields is strictly tie to the motion of the optical vortices, much of attentions have also been focused on the dynamic behavior of optical vortices in speckles. Different statistics of optical vortex motion, including the velocity distribution, the velocity variance and the trail lengths, have been examined \cite{Berry2000,Zhang2007,ZhangS2010,Alperin2019}. The velocity statistics has been proposed to serve as a measure of wave localization in strong scattering random media \cite{Zhang2007}. The persistence and lifelong fidelity related to the creation and annihilation of optical vortices show the spectrum correlation properties of the random optical fields \cite{DeAngelis2017}. Optical vortex motion analysis may also find applications in investigating the dynamics of light-scattering suspensions \cite{Kirkpatrick2012}, obtaining nanometric measurements of displacements \cite{Wang2006,Sendra2009}, and tracking cellular movement and microcirculation \cite{Wang2008,Bender2021,Gong2022}. Despite the primary importance of optical vortex stochastic motion in fundamentals and applications, the insights of optical vortex collective motion in random wavefields remains in paucity. By noticing that the discrete phase singularities perform particle-like random motion in stochastic wavefields, the vortex motion was described as the particle diffusion \cite{Cheng2014,DeAngelis2021}. However, only the pure Brownian diffusion of optical vortices have been reported \cite{OHolleran2008,OHolleran2009,Cheng2014}, and the anomalous diffusion of optical vortices has not been demonstrated yet, let alone the underlying physical mechanism.

Here we report our experimental observations of the anomalous diffusion of the optical vortices in random wave fields and reveal the underlying mechanisms of the anomalous diffusion of the optical vortices. We first follow the motion of each optical vortex by tracking the trajectories of the optical vortices from creation to annihilation and demonstrate that the optical vortices perform a typical subdiffusive behavior, the nonlinear temporal increase of the mean squared displacement (MSD), $ \left\langle \Delta r^2(\tau)\right\rangle \sim \tau^\alpha $, with $ 0<\alpha<1 $. We next model the vortex subdiffusive behavior with the fractional Brownian motion (FBM), a Gaussian process commonly used to model the particle diffusion in a viscoelastic environment. Most surprisingly, we observe a robust non-Gaussian behavior in the probability distribution of the vortex displacements. Moreover, we tune the degree of the subdiffusion and the Gaussianity of the optical vortex motion by modifying the viscoelasticity of the sample, suggesting that the optical vortices may serve as surrogate probes in optical micro-rheology for turbid media \cite{Kirkpatrick2012,Gong2022}. The knowledge obtained from this study may not only enrich our understanding of the optical vortex diffusion in random wavefields, but also contribute to the further investigations of a variety of fundamental physical phenomena, including wave transport in disordered media \cite{Gateau2019}, turbulent diffusion in quantum fluids \cite{Carusotto2013,Alperin2019}, and the riddle of anomalous yet non-Gaussian diffusion \cite{Metzler2017}.
\section{Methods}
\begin{figure}[h]
\centering\includegraphics{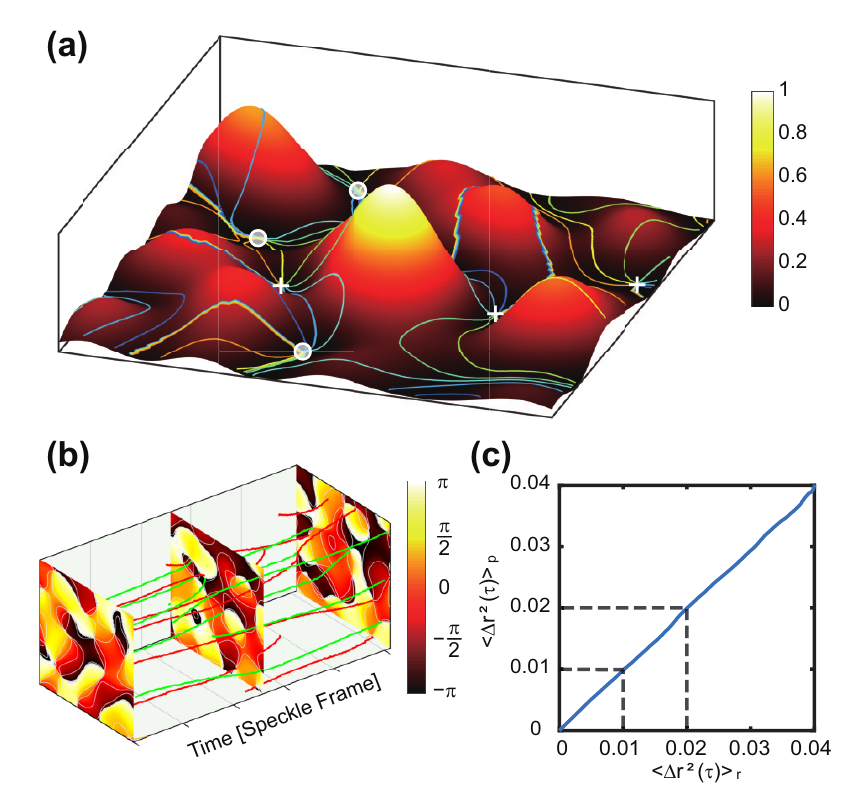}
\caption{\label{fig:Vortex}Optical vortices in speckle field. (a) Optical vortices occur at intensity nulls and the intersections of phase contours (solid lines). The phase difference is $ \pi / 4 $. The value of the intensity is represented by the color defined in the color bar. (b) The random motion of optical vortices in the generated dynamic speckle fields. The red lines and green lines represent the trajectories of the positive and negative optical vortices respectively. (c) The linear correlation between the MSD of optical vortices in the real phase and the pseudo phase.}
\end{figure}

 The collective motion of optical vortices in temporally varying random fields can be revealed by tracking the lifelong trajectories of the optical vortices \cite{DeAngelis2017}. We follow the trajectories by locating, distinguishing and linking the phase singularities in sequential speckle patterns according to the definition and the morphological features of optical vortices \cite{Kirkpatrick2012,Szatkowski2022,Wang2006}. We determine the position of a optical vortex by unwrapping the phase change along a closed counterclockwise contour around the vortex. The accumulated phase change around a single vortex is $ \pm2\pi$, corresponding to the topological charge of +1 or -1. High-order vortices are unstable and thus unlikely in random wavefields. The locations of the identified vortices over multiple frames trace a distinct vortex trail. The criterion of the new trail and existing trail is determined by the spatial correlation of speckles, or average speckle size. The vortex trajectories are then followed from creation to annihilation, as shown in Fig. \ref{fig:Vortex}(b).
 
 \subsection{Pseudo phase generation for speckle intensity}
To acquire or recover the phase of the fast-varying laser speckle fields is an arduous task due to the considerable complications of experiments and algorithms, we thus construct the pseudo phase representation of the speckle by applying the Laguerre-Gauss transformation to the speckle intensity as follows \cite{Wang2006-2}:
 \begin{eqnarray}
\Tilde{E}(x,y) & = & \left|\Tilde{A}(x, y)\right| \cdot \exp \left[j \Tilde{\varphi}(x, y)\right] \nonumber\\ & = & \iint LG\left(f_{x}, f_{y}\right) \mathcal{F}\left(f_{x}, f_{y}\right) \exp \left[j 2 \pi\left(f_{x} x+f_{y} y\right)\right] d f_{x} d f_{y},
\label{eq:LG transform}
\end{eqnarray}
where $ L G\left(f_{x}, f_{y}\right) $ is a Laguerre-Gauss transformation in the frequency domain defined as $ LG\left(f_{x}, f_{y}\right)=\left(f_{x}+j f_{y}\right) \exp \left[-\left(f_{x}^{2}+f_{y}^{2}\right) / \omega^{2}\right] $, $\mathcal{F}\left(f_{x}, f_{y}\right) $ is the Fourier spectrum of speckle intensity $I\left(x,y\right) $, $ \Tilde{E}(x, y)$ and $ \Tilde{\varphi}(x,y) $ is the generated pseudo amplitude and phase. 

Although the pseudo phase cannot reproduce the true phase of an imaged speckle pattern and the statistics properties of the optical vortices cannot be fully maintained, the spatial-temporal behavior of the optical vortices in the true phase are strictly followed \cite{Kirkpatrick2012,Veiras2016}. To verify whether the motion of the optical vortices in the pseudo phase can strictly represent that of the optical vortices in the true phase, we numerically simulate the optical vortices moving in the fluctuating speckle fields. We first generate the evolving speckle fields with different decorrelation rates by superposing a large number of partial waves from random moving scattering centers, more details are given in Supplement 1, Section 2. The complex representations of the simulated speckle intensity patterns are calculated according to Eq. \ref{eq:LG transform}. We then follow the vortex trajectories in both the pseudo and the true phase and calculate the MSD, respectively,
\begin{eqnarray}
\left\langle\Delta r^{2}(\tau)\right\rangle=\left\langle|\vec{r}(t+\tau)-\vec{r}(t)|^{2}\right\rangle,
\end{eqnarray}
where $ \vec{r}(t) $ and $ \vec{r}(t+\tau) $ are the locations of the optical vortices at times $t$ and $ t+\tau$, $ <>$ represents the ensemble average for all optical vortices. The displacement of optical vortex is normalized by the speckle size. We compare the MSD of the optical vortices in the true phase, denoted as $\left\langle\Delta r^{2}(\tau)\right\rangle_{r} $, and the MSD of optical vortices in the corresponding pseudo phase, $\left\langle\Delta r^{2}(\tau)\right\rangle_{p} $ for all simulated speckle sequences and observe an excellent equivalence between the normalized $\left\langle\Delta r^{2}(\tau)\right\rangle_{r} $ and $\left\langle\Delta r^{2}(\tau)\right\rangle_{p} $ as shown in Fig. \ref{fig:Vortex}(c). It clearly demonstrates that the averaged displacements of the optical vortices in the true phase and the pseudo phase of the sequential speckle patterns are the same. We further demonstrate that the velocity statistics of the vortices in the true phase and the pseudo phase remain consistent in Supplement 1, Section 2. Hence we can explore the motion of the optical vortex in the pseudo phase rather than in the true phase of dynamic speckle sequences.

\subsection{Experimental setup and sample preparation}
The optical setup of our experimental system is shown in Supplement 1. To experimentally generate random wavefields with optical vortices, we focus He-Ne laser ($\lambda = 632.8 nm $) into a dynamic random sample, composed of light scattering particles, Titanium Dioxide ($\mathrm{TiO_2}$) microspheres, and optical clear surrounding media, a Polydimethylsiloxane (PDMS) gel. Light scatterers, TiO$ _{\text{2}} $ nanoparticles (dia. $ \sim $500nm), are randomly distributed in PDMS matrix in the weight ratio of 1:48. PDMS (Sylgard 184, Dow Inc.) is prepared by mixing the cross-linking agent and base elastomer in the ratio of 1:10, the resulted sample reduced scattering coefficient is calculated $ \mu _{s} '=10 \ cm^{-1} $. The entire process of PDMS gelation is maintained at 37 $ ^{\text{o}}\text{C} $. The rapid-fluctuating speckle patterns reflected from the sample are captured by a high-speed CMOS camera (acA2000-340cm, Basler AG.) at the frame rate of 300 fps. We collect 13 dynamic speckle sequences with 800 patterns each at intervals of 30 minutes during the process of PDMS sample gelation.

\subsection{Spatial distribution of optical vortices}
\begin{figure}[b]
\centering\includegraphics[width = 13.2cm]{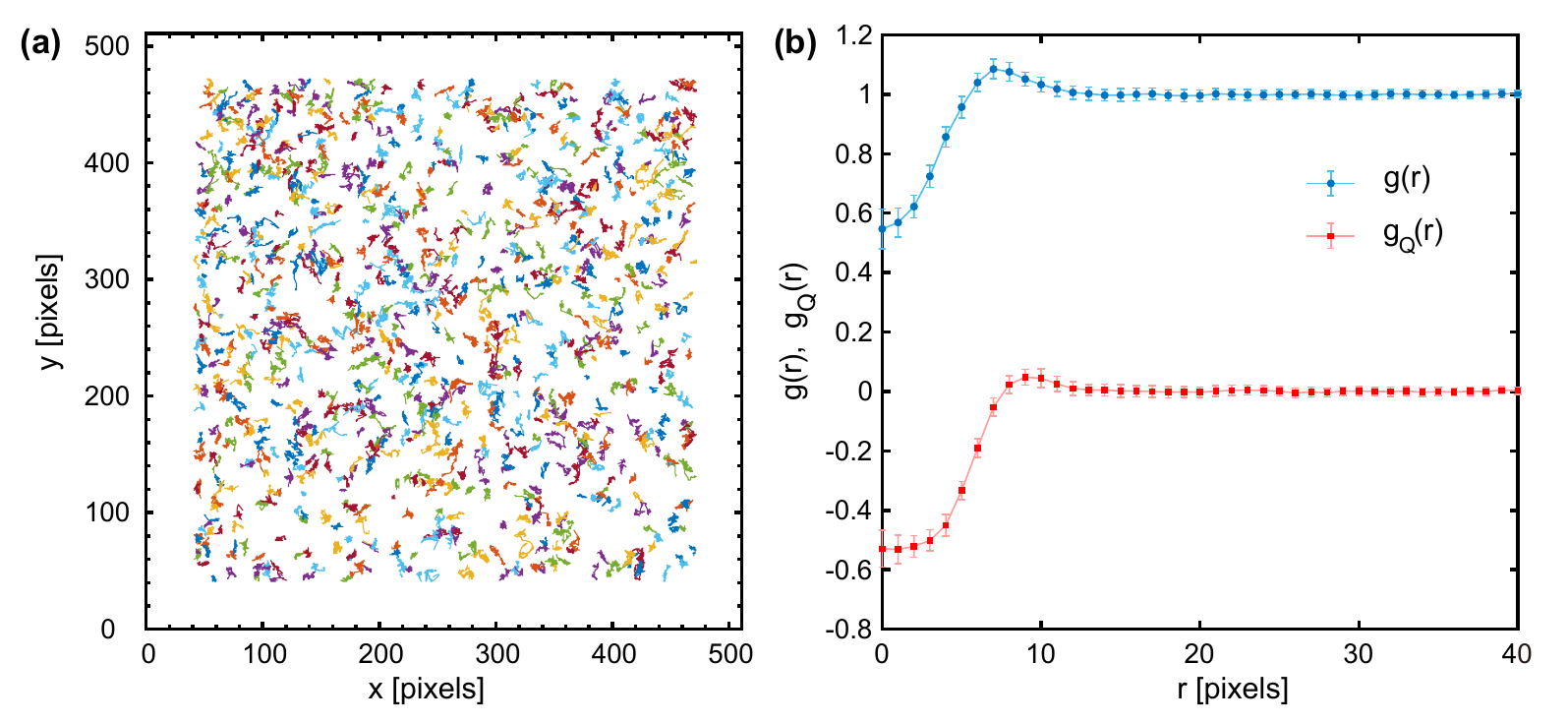}
\caption{\label{fig:Spatial Distribution}The spatial distribution of optical vortices. (a) The trajectories of the optical vortex in the pseudo phase from dynamic speckle patterns. It is clear that the optical vortex motion yields a stochastic process. (b) Pair (indigo circles) and charge (red squares) correlation function of optical vortices as a function of distance $r$. The error bars correspond to the standard deviations over 200 frames.}
\end{figure}
The fully-developed Gaussian random fields are manifested by the Rayleigh distribution of the speckle intensity shown in Supplement 1, Section 1. We investigate the two-dimensional spatial distribution of the optical vortices in the pseudo phase of the recorded speckle pattern. As can be seen in the Fig. \ref{fig:Spatial Distribution}, optical vortex trajectories are randomly and uniformly distributed in the entire speckle pattern, and individual optical vortex travels a random walk in finite space. The speckle patterns captured at the initial state of the PDMS curing process decorrelate rapidly, thus the optical vortices possess a considerable proportion of short trails. Here, for simplicity, we only show the trails of optical vortices which can survive over 150 continuous frames. Fig.\ref{fig:Spatial Distribution}(a) adequately and clearly demonstrate the random spatial-temporal distribution and the random motion of the optical vortices. We quantitatively characterize the spatial correlation of optical vortices by calculating the pair and charge correlation functions \cite{DeAngelis2016,DeAngelis2019}, 
\begin{eqnarray}
g(r) & = & \frac{1}{N \rho}<\sum_{i \neq j} \delta\left(r-\left|\boldsymbol{r}_{i}-\boldsymbol{r}_{j}\right|\right)>\\
g_{Q}(r) & = & \frac{1}{N \rho}<\sum_{i \neq j} \delta\left(r-\left|\boldsymbol{r}_{i}-\boldsymbol{r}_{j}\right|\right) q_{i} q_{j}>
\label{eq:charge correlation}
\end{eqnarray}
where $N$ is the total number of vortices, $\rho$ is the mean density, and $\delta$ is the Dirac function. Figure \ref{fig:Spatial Distribution}(b) shows the correlation statistics $ g(r)$ and $ g_{Q}(r)$ for the optical vortices in the pseudo phase of speckles. The pair correlation function $ g(r)$ return to unity as the distance $r$ increases, indicating the vortices are spatially independent for $ r \to \infty $. When $ r $  approaches to 0, $ g(r) $ is a finite positive value, meaning that we may find two vortices at extremely close position, for instance, two neighboring vortices with opposite charge. This property has been applied to super-resolution imaging \cite{Bender2021}. Meanwhile, charge correlation $ g_{Q}(r) $ is almost equivalent to $ -g(r) $ for $r \to 0$. These observations reflect representative repulsion between the optical vortices, for instance, two neighboring vortices with same charge. All the observed results are completely consistent with the characterizations of optical vortices in isotropic random waves \cite{Berry2000}.

\section{Results and Discussion}

\subsection{Subdiffusion of the optical vortices}

\begin{figure}[b]
\centering\includegraphics{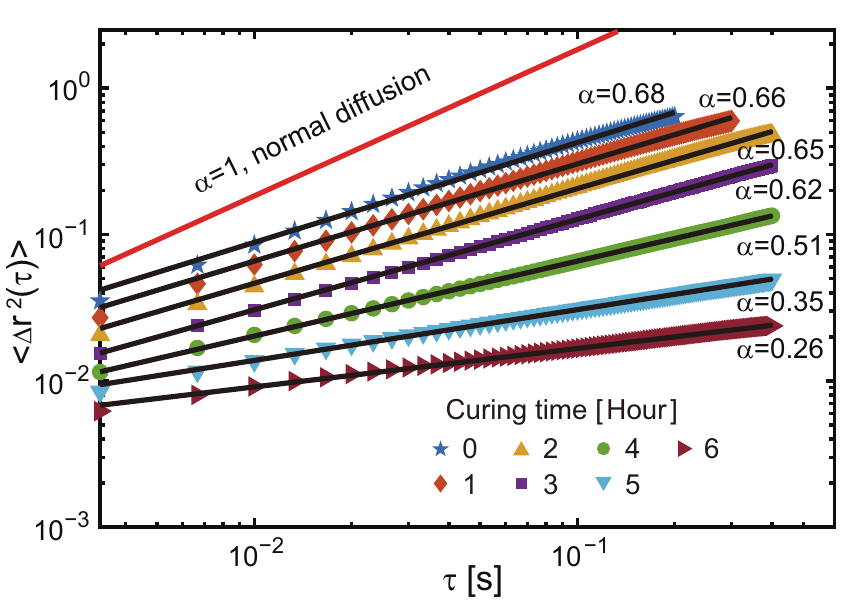}
\caption{\label{fig:Subdiffusion}Anomalous diffusion of optical vortices. Double-logarithmic plot of the MSD for optical vortices in dynamic speckles recorded at different curing time during the PDMS gelation. The data are averaged over 30,000 vortex trajectories with different trail lengths. The black solid lines represent the fits of the MSDs to the power-law relation $ \left\langle \Delta r^2(\tau)\right\rangle \sim \tau^\alpha $, and the corresponding $ \alpha $ are listed. The red solid line depicts a normal diffusion.}
\end{figure}
 The stochastic motion are commonly described as Brownian diffusion with the mean squared displacement increasing linearly with time. Deviations from such a linearity termed as anomalous diffusion are also common in transport dynamics from quantum physics to life sciences \cite{Aniello2019,Hofling2013}. Figure \ref{fig:Subdiffusion} shows the MSDs of optical vortices with the time lag $\tau$ at different curing times during the PDMS gelation. Throughout the whole PDMS gelation process, all the MSDs increase nonlinearly with time lag $\tau$ and demonstrate the power-law dependence of $\tau$, $ \left\langle \Delta r^2(\tau)\right\rangle \sim \tau^\alpha $. The time lag $\tau$ spans nearly three decades for most MSDs. The scaling exponent $\alpha$ for all MSDs are always less than unity, indicating that the optical vortices performs the subdiffusive motion in the Gaussian random wavefields. At the early curing times, due to the low viscosity of the PDMS sample, a rapid decorrelation of the speckle patterns is observed as shown in the Supplementary Materials. The corresponding MSD of the optical vortices increases rapidly suggesting a rapid diffusion of optical vortices. The exponent $\alpha$ decreases from 0.68 to 0.26 with the stiffening of the viscoelastic PDMS sample. For the almost cured PDMS sample, we only observe a slight increment in the MSD curves, which suggests the vortex motion is tightly restricted. The subdiffusive behavior of optical vortices and the constrained motion with the viscoelasticity resemble those of particles in viscoelastic media. Single or multiple particle tracking microrheology is a well-established micro-rheological technology which measures rheological properties of soft matter by tracking the stochastic motion of probe particles embedded in the sample. Our results directly propose a novel optical micro-rheology which track the optical vortices rather than the exogenous probe particles to evaluate the viscoelasticity of turbid media.
 
\subsection{Underlying mechanism of the subdiffusive behavior}
\begin{figure}[b]
\centering\includegraphics[width = 13.2cm]{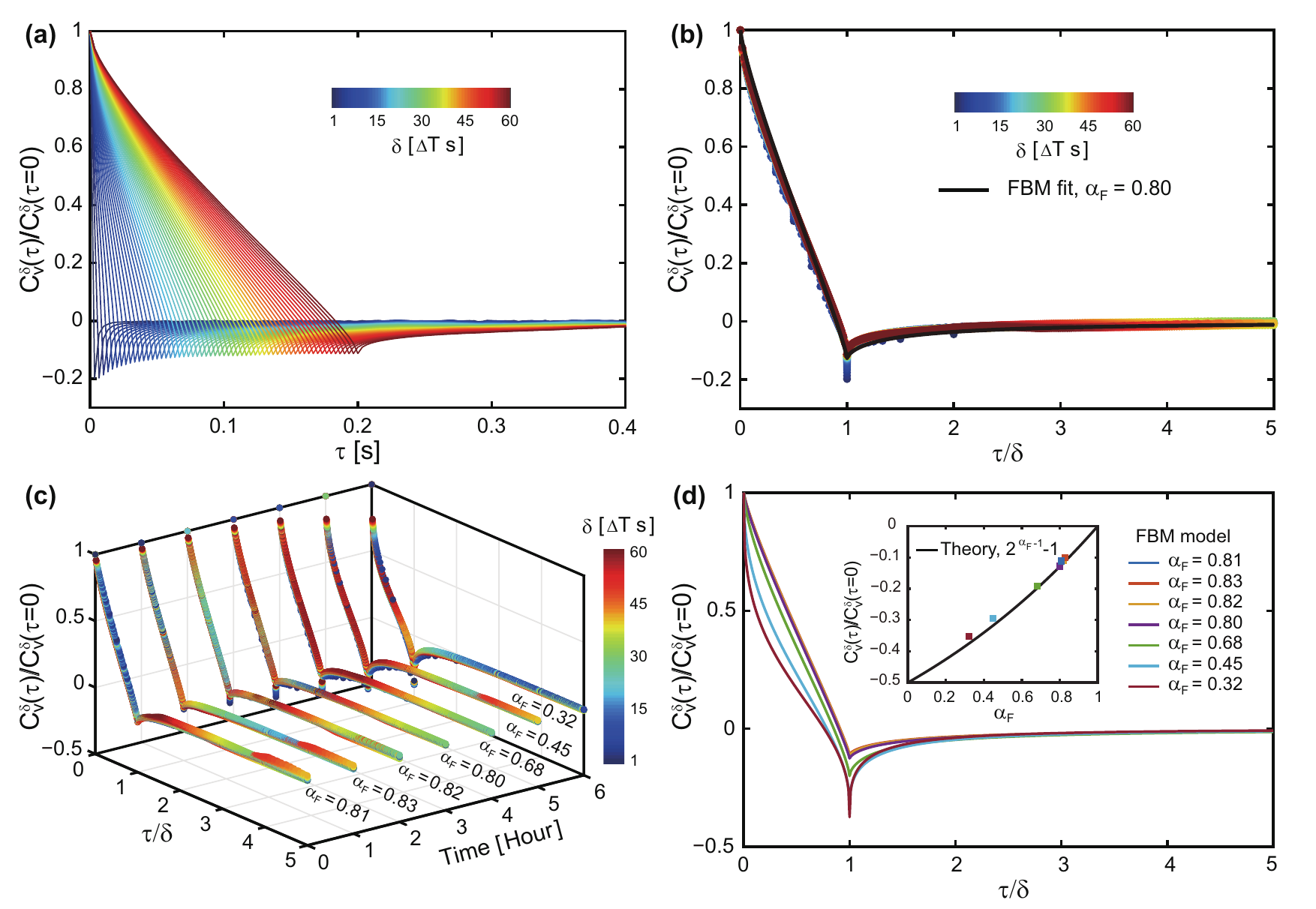}
\caption{\label{fig:VelocityCorrelation}Velocity autocorrelation function $ C_{v}^{\delta}(\tau) / C_{v}^{\delta}(\tau=0) $ for the anomalous diffusion of optical vortices. The average velocity is calculated at $ \delta=n \times \Delta T s \left( n=1,2,3,...,60\right)  $ (blue to red). $ C_{v}^{\delta}(\tau) / C_{v}^{\delta}(\tau=0) $ is plotted against (a) the time lag $ \tau $ and (b) the rescaled time lag $ \tau / \delta $. Data are obtained at the time point of 3 hour. The solid black line is the FBM fit of the data. (c) The $ C_{v}^{\delta}(\tau) / C_{v}^{\delta}(\tau=0) $ curves acquired at different curing time points versus the rescaled time lag $ \tau / \delta $. (d) The corresponding FBM fit of the collapsed curves in Fig. 4(c). The scaling exponent $ \alpha_\mathrm{F} $ ranges from 0.81 to 0.32 as the PDMS sample curing. Inset: the minimum value of the VAF at $ \tau / \delta $ as a function of scaling exponent $ \alpha_\mathrm{F} $. The solid line is the theoretical prediction with the FLM model.}
\end{figure}
Subdiffusive motion could be induced by several different mechanisms. For example, the subdiffusion of particles in a viscoelastic medium can be modeled by fractional Brownian motion model, possessing coherent increments in the particle's motion \cite{Mandelbrot1968}. Continuous time random walk (CTRW) model describes a general process dictated by a sequence of binding–unbinding events in crowded environment \cite{Metzler2000}. To further the investigation of the subdiffusive behavior of the optical vortices in random wavefields, we calculate a diagnostic function \cite{Weber2010,Weber2012}, the velocity autocorrelation function:
\begin{eqnarray}
C_{v}^{\delta}(\tau)=\langle\vec{v}(t+\tau) \cdot \vec{v}(t)\rangle,
\label{eq:velocity autocorrelation}
\end{eqnarray}
where the averaged velocity $\vec{v}(t)=\frac{1}{\delta}[\vec{r}(t+\delta)-\vec{r}(t)]$, the time interval $ \delta =n\times\Delta T $ s $ (n=1,2,3...,60) $ and $\Delta T $ is the smallest increments of time in experiments. Fig. \ref{fig:VelocityCorrelation}(a) shows the VAFs of the optical vortex motion in random speckles recorded at 3 hours after the curing of PDMS started. Each VAF curve in Fig. \ref{fig:VelocityCorrelation}(a) has different time interval $\delta$. We observe all VAF curves reach to a dip into negative values at time $\delta$, indicating the negative correlations of the optical vortex motion in intermediate time. The negative correlation depictes a push-back tendency of the optical vortices in the speckle which may result from the anticorrelation between the vortices \cite{Shvartsman1994}. The observed antipersistent correlation is the hallmark of FBM. We also notice that the VAF curves rescaled by the time of the negatives dips collapse to a universal curve as shown in Fig. \ref{fig:VelocityCorrelation}(b). The universal curve of VAF reveals the self-similarity of the optical vortex motion, similar patterns at different temporal and spatial scales. The self-similar and negatively correlated behaviors of the vortex motion also exclude other possible mechanisms of the subdiffusion including localizing errors in tracking optical vortex, and CTRW \cite{Weber2010,Weber2012}.

Furthermore we fit the measured velocity autocorrelation function with the theoretical prediction of VAF in FBM model as \cite{Weber2010,Burov2011}
\begin{eqnarray}
C_{v}^{\delta}(\tau) / C_{v}^{\delta}(\tau=0)=\frac{(\eta+1)^{\alpha_\mathrm{F}}+|1-\eta|^{\alpha_\mathrm{F}}-2}{2 \eta^{\alpha_\mathrm{F}}},
\label{eq:rescaled-VAF}
\end{eqnarray}
where $1/\eta $ is the rescaled time lag $ \tau/\delta$. The rescaled VAF shows an excellent agreement with Eq. \ref{eq:rescaled-VAF}, with $\alpha_\mathrm{F}$ as the only fitting parameter, as shown in Fig. \ref{fig:VelocityCorrelation}(b). It indicates that the subdiffusive behavior of optical vortices can be well described by the FBM model. Additionally, we calculate the VAF for optical vortices in speckle sequences obtained at different curing time points throughout the PDMS curing process. As shown in Fig. \ref{fig:VelocityCorrelation}(c), all the rescaled VAF curves at different time points show the same tendency of collapsing onto a unique master curve respectively. We fit the collapsed curves to the FBM model and retrieve the scaling exponents $\alpha_\mathrm{F}$  correspondingly. The $\alpha_\mathrm{F}$ is propotional to the $\alpha$  retrieved from the MSD of the optical vortices. Fig. \ref{fig:VelocityCorrelation}(d) shows the analytical rescaled $C_{v}^{\delta}(\tau)$ in FBM model. As the PDMS sample gelating, the negative dip of the VAF is approaching the theoretical limit (-0.5) which suggests the extreme tight confinement. In the inset, we plot the dip values averaged over all time intervals $\delta$ at $\tau=\delta$ against scaling exponent $\alpha_\mathrm{F}$ and our experimental results show excellent agreement with the analytical expression in Eq. \ref{eq:rescaled-VAF}, $2^{\alpha_\mathrm{F}-1}-1$. This consistency further support the claimed FBM behavior of optical vortices in dynamic speckles. We conclude that the optical vortices in random wavefields are undergoing subdiffusive motion in terms of viscoelastic diffusion as FBM.

\subsection{Non-Gaussianity in FBM}
Another significant feature of the normal Brownian diffusion, besides the linear increase in time of the mean squared displacement, is the Gaussian probability distribution of the particle displacements. Brownian motion and Gaussian process were considered to be intimately tied together according to the central limit theorem. This connection was enforced by the observations that the particles in pure viscous media undergo Brownian motion accompanied with a Gaussian displacement distribution while certain particle thermal motion displays anomalous diffusion coexisting with a non-Gaussian displacement distribution \cite{Metzler2000,Hofling2013,Jeon2016,Lampo2017,Metzler2017}. However, numerous recent reports for different soft material and biological systems have demonstrated the clear decoupling between the "Brownianity" and "Gaussianity". The intriguing Brownian yet non-Gaussian diffusion has been discovered in crowded colloids \cite{Wang2009-3,Guan2014,Pastore2021}, cells and active matter \cite{Kurtuldu2011,He2016,Kanazawa2020,Lastra2021}. The Gaussian but anomalous diffusion has also been identified in dilute solutions \cite{Metzler2000}.
\begin{figure}[h]
\centering\includegraphics[width=7cm]{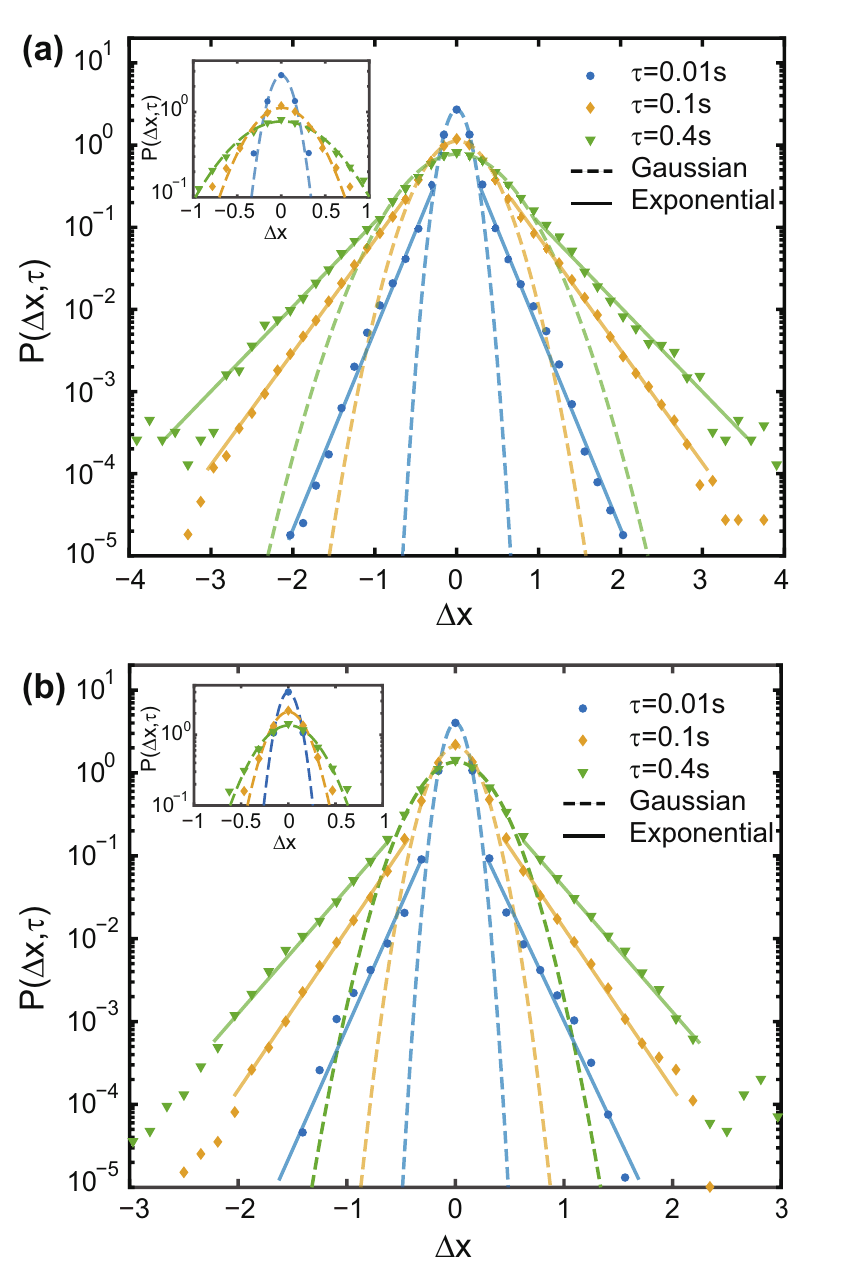}
\caption{\label{fig:Nongaussian}Non-Gaussian behavior of optical vortices. (a) Displacement PDFs over 6 orders of magnitude plotted against displacement, $ \Delta x$, at the intial state of PDMS gelation, at time lags 0.01, 0.05, 0.1, 0.2, and 0.4 s. The dashed lines represent the Gaussian distribution and the solid lines show the exponential distribution at corresponding time lags. The inset shows log-linear plot of the same data at small $ \Delta x$ values. (b) Data are obtained at PDMS curing time points of 3 hour.}
\end{figure}

Here we already demonstrate the optical vortex subdiffusive motion conforming to a Gaussian process, the FBM, suggesting that the optical vortex may perform a Gaussian anomalous diffusion. We further studied the probability distribution of optical vortex displacements and surprisingly found obvious deviations from the Gaussian distribution. As shown in Fig.\ref{fig:Nongaussian}, the PDFs of optical vortex transverse displacements $ \Delta x $ at different lag time $ \tau $ are perfectly fit to the Gaussian distributions only at small displacements. At large displacements, the PDFs deviate from the Gaussian distributions and transit to the heavy exponential tails quickly. The PDFs with the sharp transitions from Gaussian distributions to exponential tails are similar to the PDFs of the Brownian yet non-Gaussian diffusion. By comparing the Fig. \ref{fig:Nongaussian} (a) and (b), we also notice that the PDFs possess a narrower width at the later gelling stage (Fig. \ref{fig:Nongaussian}(b)) than the PDFs at an early gelling stage (Fig. \ref{fig:Nongaussian}(a)). The reason is, at the later gelling stage, the light scattering particles moves slowly in the media with higher viscoelastic modulus causing the laser speckles vary slowly and hence the optical vortices moves slowly. The underlying physical mechanism of the non-Gaussian behavior in normal or anomalous diffusion are still in highly debate \cite{Wang2012,Chubynsky2014,Metzler2017,Sposini2018,Metzler2020,Chakraborty2020}. The discovered non-Gaussian behavior of optical vortex subdiffusion may also enrich the understanding of this general phenomenon.

Vortices or phase singularities are not only present in wave fields, but also in many different physical systems, where they are better known as quantized vortices in superfluids \cite{Eyink2006,Fetter2010,Tang2021} or topological defects in liquid crystals and even in cell membranes \cite{Kleman1989,Brasselet2011,Tan2020}. The analogy between the underlying physics that govern the vortices in different systems has been well recognized for decades. For example, the wave coherence in random wave fields has been admitted to be an analogy to the quantum coherence in turbulent superfluids \cite{Carusotto2013,Alperin2019}. Hence the above analysis of optical vortex motion may also contribute to the investigation the collective dynamics of other vortices.

\section{Conclusion}
In conclusion, we present the experimental evidence of the anomalous diffusion of optical vortices in random wavefields. We found that the subdiffusive motion of optical vortices clearly show two crucial features, the self-similarity and antipersistent behavior which can only be interpreted by the fractional Brownian motion model. Most surprisingly, we observe a robust non-Gaussian behavior in the probability distribution of the optical vortex displacements, which is directly contradict to the implication of the Gaussian process of the fractional Brownian motion. In addition, we also find the extent of the subdiffusion of optical vortex can be tuned by the changes in the viscoelasticity of the light scattering media. The findings of anomalous yet non-Gaussian diffusion of optical vortices not only enrich our understanding of the relation between the Brownianity and Gaussianity, but also extend the knowledge of anomalous diffusion of vortices from systems with rest masses like fluids or superfluids \cite{Tang2021} to the massless systems like the phase or polarization of random light fields \cite{Sugic2021,Gbur2016,Chriki2019,Sitnik2022}. Due to the recognized analogy between the quantum coherence of superfluids and the wave coherence in random optical fields \cite{Carusotto2013,Alperin2019}, the analysis of optical vortex dynamics may gain insights into the dynamics of quantum turbulence, and of course vise versa. Moreover, the reminiscence between the subdiffusion of optical vortices in random optical fields from turbid viscoelastic media and the particle subdiffusion in viscoelastic environment strongly suggests an optical vortex tracking based microrheology approach, particularly for turbid soft matter or biological tissues, as a counterpart of single or multiple particle tracking microrheology.

\begin{backmatter}
\bmsection{Funding}
Science, Technology and Innovation Commission of Shenzhen Municipality (JCYJ201708181601
18360).

\bmsection{Acknowledgments}
We thank Prof. Shaoqun Zeng (Huazhong University of Science and Technology) for enlightening discussions and for commenting on our manuscript. 

\bmsection{Disclosures} 
\noindent The authors declare no conflicts of interest.

\bmsection{Data availability}  Data underlying the results presented in this paper are not publicly available at this time but may be obtained from the authors upon reasonable request.

\bmsection{Supplemental document}
See Supplement 1 for supporting content. 
\end{backmatter}


\bibliography{Manuscript}






\end{document}